\documentclass[convention,peer-reviewed]{aesconf}

\graphicspath{{./}{figures/}}

\usepackage[utf8]{inputenc}

\usepackage{microtype}

\usepackage[numbers,square]{natbib}

\usepackage{booktabs}
\usepackage{color}
\usepackage{url}

\usepackage{amsmath}
\usepackage{booktabs}

\usepackage[hidelinks]{hyperref}
\usepackage{amsfonts}
\usepackage{pbox}

\title{Advances in Thunder Sound Synthesis}

\author[1,3]{Eva Fineberg}
\author[1,2]{Jack Walters}
\author[1,2]{Joshua Reiss}

\affil[1]{Queen Mary University of London}
\affil[2]{Nemisindo Ltd}
\affil[3]{Native Instruments GmbH}

\correspondence{Eva Fineberg}{eva.fineberg@gmail.com}

\lastnames{Fineberg, Walters, and Reiss}

\shorttitle{Advances in Thunder Sound Synthesis}

\savebox{\AEStop}{%
	\begin{minipage}{\textwidth}%
		\rule{\textwidth}{1.5pt}\\%
		\\%
		\begin{minipage}[c][\iftoggle{convention}{3.2cm}{3.7cm}][t]{0\textwidth}%
			\includegraphics[width=20mm]{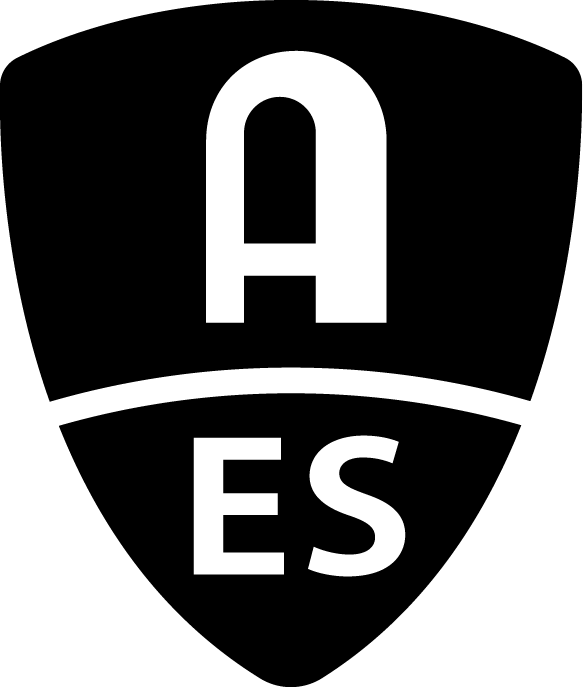}%
		\end{minipage}%
		\begin{minipage}{\textwidth}%
			\sffamily%
			\begin{center}%
				\LARGE Audio Engineering Society\\%
				\iftoggle{e_brief}{%
				\hspace{3mm}\fontsize{36}{38pt}\selectfont Convention e-Brief \AESEBriefNumber\\%
				}{%
				\iftoggle{convention}{%
				\fontsize{36}{38pt}\selectfont Convention Paper\\%
				}{%
				\fontsize{36}{38pt}\selectfont Conference Paper\\%
				}}%
				\vspace{0.2cm}%
				\large Presented at the \AESConferenceNumber \iftoggle{convention}{Convention\\}{\AESConferencePrefix Conference on\\}%
				\iftoggle{convention}{}{\AESConferenceTopic\\}%
				\AESConferenceDate, \AESConferenceLocation%
			\end{center}%
		\end{minipage}\\%
		\vspace{0.2cm}\\%
		\begin{minipage}{\textwidth}%
			\rmfamily\itshape\small	\AESLegalText%
		\end{minipage}\\%
		\\%
		\rule{\textwidth}{1.5pt}%
	\end{minipage}%
}

\begin{document}

\twocolumn[
\maketitle % MANDATORY!

\begin{onecolabstract}
A recent comparative study evaluated all known thunder synthesis techniques in terms of their perceptual realness. The findings concluded that none of the synthesised audio extracts seemed as realistic as the genuine phenomenon. The work presented herein is motivated by those findings, and attempts to create a synthesised sound effect of thunder indistinguishable from a real recording. The technique supplements an existing implementation with physics-inspired, signal-based design elements intended to simulate environmental occurrences. In a listening test conducted with over 50 participants, this new implementation was perceived as the most realistic synthesised sound, though still distinguishable from a real recording. Further improvements to the model, based on insights from the listening test, were also implemented and described herein.
\end{onecolabstract}
]

\medskip

\section{Introduction}
\label{ref:introduction}
As the field of procedural audio advances, generative models produce increasingly realistic, flexible sound effects, prompting an industry shift towards the use of procedural audio in favour of sound samples. Generative synthesised sound effects can dynamically alter context-dependent mixing techniques and audio effects, affording the sound designer broader creative agency and control. This is explored in a small canon of work focused on audio models which produce a ``physically accurate simulation with designer control'' \cite{glassner_2000}.

\bigskip

This paper aims to grow this collection of work and contribute a high-quality, flexible, thunder synthesis model indistinguishable from the sound of the real event. Based on FXive's implementation, our model's novel components hone the translation of physics-based sonic occurrences into design components \cite{fxive}. The model is real-time and implemented in the browser using The Web Audio API \cite{wc3workinggroup}. The subjective evaluation compares this model with select existing synthesis models as well as a real recording, and the results provide a detailed profile of our success.

This paper is organised as follows. Section \ref{ref:relatedwork} summarises related work in the field and Section \ref{ref:themodel} introduces our proposed model for thunder synthesis. Section \ref{ref:evaluation} describes the evaluation method, the results of which are presented in Section \ref{ref:results} and discussed in Section \ref{ref:discussion}, before we provide our concluding remarks in Section \ref{ref:concludingremarks}.
% \enlargethispage{14pt}

\section{Related Work}
\label{ref:relatedwork}

Research exploring the sound patterns which comprise thunder has demonstrated these patterns are difficult to model exhaustively \cite{ribner_roy_1982}, \cite{glassner_2000}. In their comparative study, Tez, Selfridge, and Reiss examined nine published and unpublished models that attempt to synthesise the sound of thunder \cite{tez_selfridge_reiss}. These models can be broadly classified into physics-based and signal-based models. The former aims to recreate the physical environment in which the sound effect occurs, whereas the latter uses procedural design principles to imitate the perceived sound patterns. 
\begin{figure*}
\centering
    \includegraphics[scale=0.1]{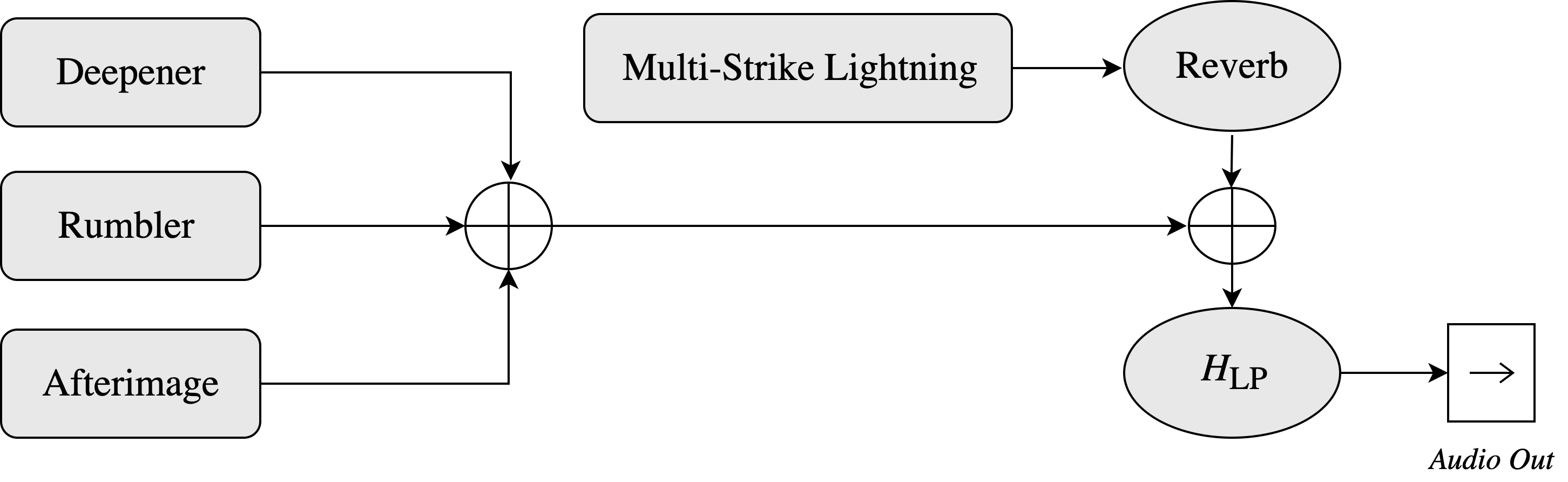}
\caption{High level block diagram of the complete thunder synthesis model.}
\label{fig:lightningoverall}
\end{figure*}

\subsection{Physics-based Models}
\label{ref:subrelatedwork-physical}

One of the more notable contributions to this field is the work of Ribner and Roy \cite{ribner_roy_1982}. They detail an approximate, quasilinear theory of thunder generation physics whereby an expanding cylindrical shockwave unfolds and emanates from points along a segment of a tortuous lightning channel \cite{ribner_roy_1982}. 

At its core, \cite{ribner_roy_1982} first constructs the lightning segments from which the channel is composed, then amalgamates the net effect of the N-shaped pressure profiles originating from those segments. The N-wave, depicted in Fig. 2 in \cite{wm_68}, evolves to create a time-pressure mapping of the shape of lightning onto the resulting pressure disturbances which compose the sound of thunder \cite{wm_68}. 

Wright-Mendendorp later evaluate this pressure profile and alternatively consider the linear superposition of each wave from a source. This simplified representation results in a slightly distorted shape of an N-Wave known by the authors' initials as the WM-wave. The WM-wave, depicted in Fig. 6 in \cite{wm_68}, exhibits a similarly sharp increase in pressure, and softer tail shock \cite{wm_68}. These pressure profiles have since become key in subsequently improved models such as \cite{glassner_2000} and \cite{blanco_2008}.

One physics-based approach which differs significantly from the aforementioned is Saksela's \emph{Thunder simulation} \cite{saksela_2014}. This model constructs the dissipating shockwaves using a Brode pulse as the pressure profile. Similar to WM-waves, Brode pulses evolve over time from segments along a lightning channel. However, the shape of the wave differs in that the sharper wave front reaches the listener first.

\subsection{Signal-Based Models}
\label{ref:subrelatedwork-signal}

One limitation incurred when generating physics-based models is the computational complexity of the underlying physics. This level of complexity can strain resources and impact real-time latency. A signal-based, physics-inspired approach can greatly improve this shortcoming, making itself more accessible to real-world use cases.

Farnell offers such an approach in \cite{farnell_2010} which begins with an initial multi-strike sound, triggered by a series of impulses. It is followed by a dampened N-wave-inspired noise pattern and enhanced by an `Afterimage' of environmental echoes. The long tail of the sound effect is composed of low frequencies and smaller amplitude peaks. Sound sources are influenced by each other and randomness in the system, this noticeably impacts the lasting `Rumble' and `Afterimage' sub-models which reflect environmental obstacles simulated by reflection, diffusion, and refraction. Additionally, delay and echoes further spatialise to the model \cite{farnell_2010}.

The architecture provided in \cite{farnell_2010} was implemented as a browser-based solution in \cite{fxive} and serves as the foundation of the approach herein.

\section{The Model}
\label{ref:themodel}
Studies investigating synthesised models of natural occurrences such as \cite{lee2020real-time} and \cite{unpublishedSelfridge} have found physics-based implementations were perceived as more realistic than signal based techniques. However, in the case of the sound of thunder, the findings in \cite{tez_selfridge_reiss}, found the reverse to be true. It is for this reason we propose a signal-based, physics-inspired, sound design model of thunder synthesis. There are four key sonic events which comprise the synthesised experience:

\vspace{0.1cm}

\begin{enumerate}
     \vspace{-0.2cm}\item Multi-Strike Lightning
     \vspace{-0.2cm}\item Rumbler
     \vspace{-0.2cm} \item Afterimage
     \vspace{-0.2cm}\item Deepener
\end{enumerate}

The implementation of each of these events begins with a sound source, and uses synthesis techniques influenced by user input and environmental factors to design the output spectrum.\footnote{\href{https://github.com/bineferg/thunder-synthesis}{https://github.com/bineferg/thunder-synthesis}} An overview of the model is shown in Fig.\ref{fig:lightningoverall}. High-pass ($H_{\text{HP}}$), low-pass ($H_{\text{LP}}$) and band-pass ($H_{\text{BP}}$) filters specified in \cite{wc3workinggroup}, \cite{reissworking} are used to alter the frequency response of the generated signals throughout. The filters are defined as recursive linear biquadratic filters centered around a frequency $f$ with quality factor $Q$. Audio effects such as delay, panning, and reverb enhance the filtered signals.

The complete model exists as a published web application\footnote{\href{https://nemisindo.com/models/thunder.html}{https://nemisindo.com/models/thunder.html}} and is controlled by four user-defined parameters including `distance' from the point of the initial strike sound $P_{\text{distance}}$, the intensity of the `initial strike' $P_{\text{\:initialStrike}}$, and `rumble' and `growl', denoted by $P_{\text{rumble}}$ and $P_{\text{growl}}$, respectively. The reverb button applies convolutional reverb to sub-model \ref{ref:themodel-lightningstrikes}. A detailed description of the post-processing effects used in this model are outlined in Section \ref{ref:postprocessing}.

\subsection{Multi-Strike Lightning}
\label{ref:themodel-lightningstrikes}

\begin{figure}[h]
\centering
    \includegraphics[scale=0.09]{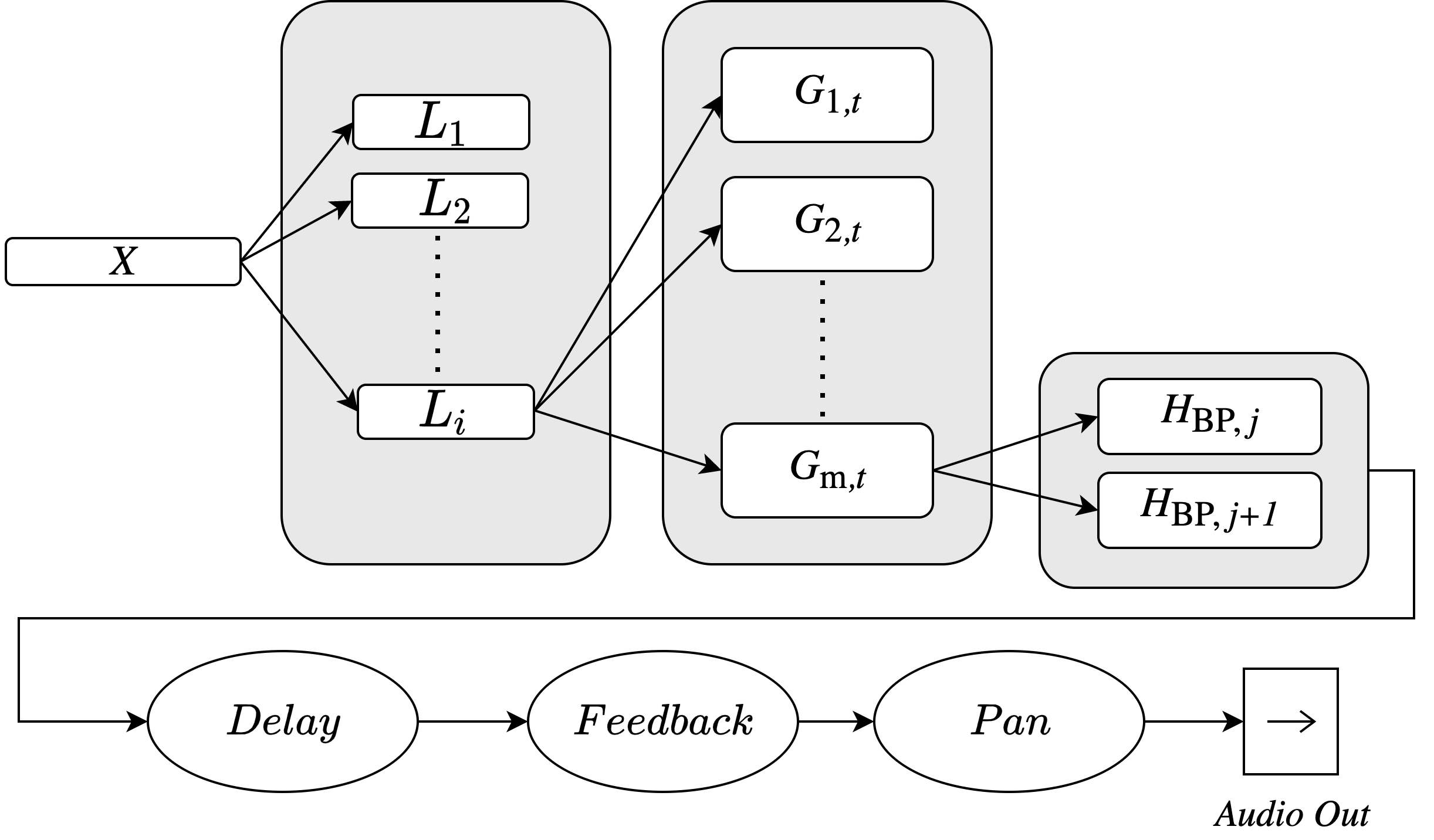}
\caption{Block diagram of the ‘Multi-Strike Lightning’ generator. A minimum of 1 strike is always generated because 62\% of lightning flashes were found to consist of 1 to 2 \emph{claps} \cite{bodhika}.}
\label{fig:singleStrike}
\end{figure}

The sonic \emph{clap} resulting from lightning strikes is perhaps the most distinctive element of the experience. Fig. \ref{fig:singleStrike} details the signal flow emulating the first acoustic shockwaves resulting from a lightning strike \cite{Graneau_1989}. The model generates $i$ lightning strikes $L_i$, and discrete random variable $i$ is distributed uniformly according to $U[1,6)$.

\begin{subequations}
    \begin{align}
    &X = 
        \begin{cases}
        WN, & \text{ if } i \: \text{ mod } \: 2 = 0\\
        \delta, & \text{ otherwise }
        \end{cases} \label{eq:strike-a}
    \\
    &L_i = H_{\text{BP},[j,j+1]}(X) \label{eq:strike-b}
    \end{align}
\end{subequations}

The filter bank $H_{\text{BP}, j}$ consists of $j+1$ time-varying band-pass filters with $Q=10$. Intermediate signal $X$ is either a white noise audio signal $WN \sim U[-1,1)$, sampled from a continuous uniform distribution $U$, or a collection of 20 impulses $\delta$

\begin{equation}
    \delta_t = 
    \begin{cases}
        1, & \text{if } t \leq  t_r + \epsilon \\
        0, & \text{otherwise}
    \end{cases}
\label{eq:strikedelta}
\end{equation}

where $t$ is time in seconds and $t_r$ is a continuous random variable distributed according to $U[0,1)$. A small $\epsilon$ is used to ensure $t_r > 0$.

\vspace{0.2cm}

$m=4$ strike envelopes controlled through a gain $G_{m,t}$ at time $t$ are used to separately envelope each $L_i$. $G_{m,t}$ is controlled by a linear ramping function $R$ over period $T=[d,d']$. As such, $R(G_{m,t})$: $P_{\text{\:initialStrike}}\cdot2 \xrightarrow[T]{} 0$, $t \in T$ and distance delay time $d=P_{\text{distance}} \cdot \frac{1}{C}$ where $C$ is the speed of sound. $d'$ is the upper bound of $T$ s.t. $d' := d + 240(1.4-r)^5$ seconds where $r \sim U(0,1)$ is a continuous random variable. As a result of this, $d'$ is bounded in $(2.45, 1290)$ms. Calculating $d'$ from a fifth power ensures its bounds are sufficiently wide and its behaviour is sufficiently non-linear.

\vspace{0.2cm}

$r$ is also used to set the cutoff frequency $f_{j,t}$ of filter $H_{\text{BP}, j}$ where $f_{j,t} = (r \cdot 1200 + 100)$Hz, and ramps linearly s.t. $R(f_{j,t})$: $f_{j,t} \xrightarrow[T]{} \frac{f_{j,t}}{2}$. \eqref{eq:strike-b} details how $X$ is filtered by $H_{\text{BP}, j}$ and $H_{\text{BP}, j+1}$, where $j=m\cdot2$. 

Strike generation may also stochastically trigger a lightning `split' process, in which a second `branch’ of lightning is generated and their processed envelopes are superpositioned. Regardless, each $L_i$ is convolved with an impulse response of a beach adding a natural reverb to the resulting signal \cite{warren_2013}.

\subsection{Rumbler}
\label{ref:themodel-rumbleandroll}
\begin{figure}[h]
\centering
    \includegraphics[scale=0.11]{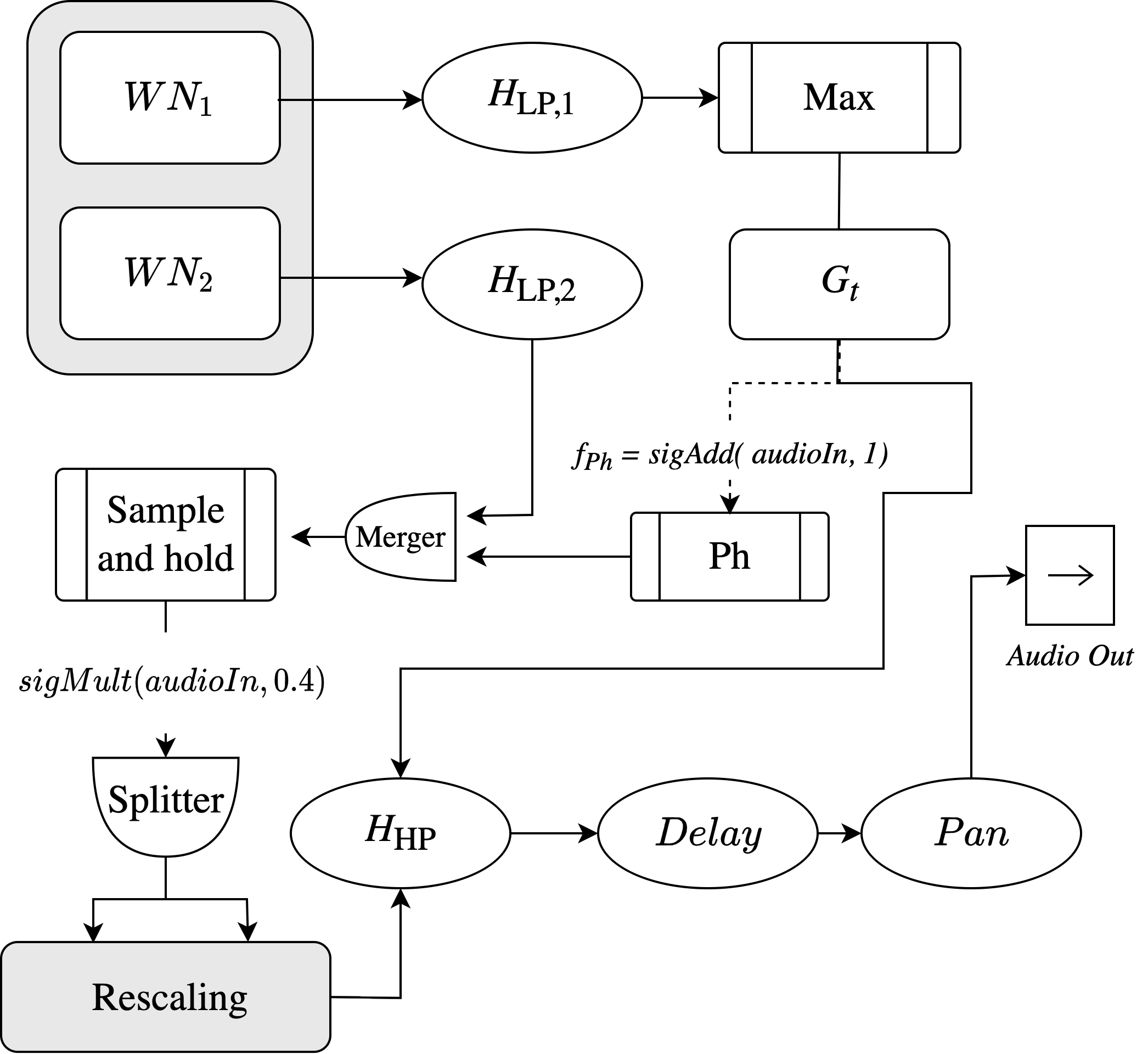}
\caption{Block diagram of a the `Rumbler' generator. The base implementation is provided by \cite{fxive}.}
\label{fig:rumbler}
\end{figure}

The rumble of thunder often precedes and follows the initial \emph{clap}, but can be experienced most prominently during the long tail of the sonic experience. Composed of varying amounts of low level frequency content, Fig. \ref{fig:rumbler} outlines the `Rumbler' signal generator \cite{bodhika}. This sub-model processes two independent white noise audio signals $WN_{[1,2]} \sim U[-1,1)$. Gain $G_{t}$ of source $WN_1$ ramps periodically according to a non-linear ramping function $R'$ s.t. $R'(G_{t})$: $P_{\text{rumble}} \cdot 2.5 \xrightarrow[T]{} 0 + \epsilon$. In this sub-model, $d' := d+9s$, and $\epsilon$ is a small value to ensure $G_{t=d'} > 0$. $R'$ achieves the effect of modelling an undulating and slowly dissipating `rumble' sound. Signal $WN_1$ is low-pass filtered by $H_{\text{LP,1}}$ and clipped to have a minimum value of 0,

\begin{equation}
    RN_1 = \text{max}(H_{\text{LP,1}}(WN_1),0)\\
    \label{eq:rumbleNoise1}
\end{equation}

resulting in rumble noise $RN_1$, where `Max' represents a max-value operation. $G_{t} + 1$ is subsequently used to set the frequency $f_{Ph}$ of a phasor $Ph$. $WN_2$ is low-pass filtered by $H_{\text{LP,2}}$ and sampled by a sample-and-hold processor evaluated as

\begin{equation}
    RN_2[n] = 
    \begin{cases}
        H_{\text{LP,2}}(WN_2)[n], & \text{if } tr[n] < tr[n-1] \\
        RN_2[n-1], & \text{if } tr[n] \geq tr[n-1]
    \end{cases}
\label{eq:sampler}
\end{equation}

to ensure that $RN_2$ is sampled and held as a function of $G_{t}$, where $n$ is the sample number and the trigger $tr[n]$ is controlled by $Ph$. Over period $T$, where $d' := d+12s$, the cutoff frequency $f_t$ of $H_{LP,1}$ and $H_{LP,2}$ are linearly ramped s.t. $R(f_t)$: 1000Hz $\xrightarrow[T]{}$ 0Hz. The signal $RN_2$ is then split, processed separately and recombined, where it is scaled as a function of itself to imitate the interconnected nature of thunder physics.

\subsection{Afterimage}
\label{ref:themodel-Afterimage}

\begin{figure}[h]
\centering
    \includegraphics[scale=0.095]{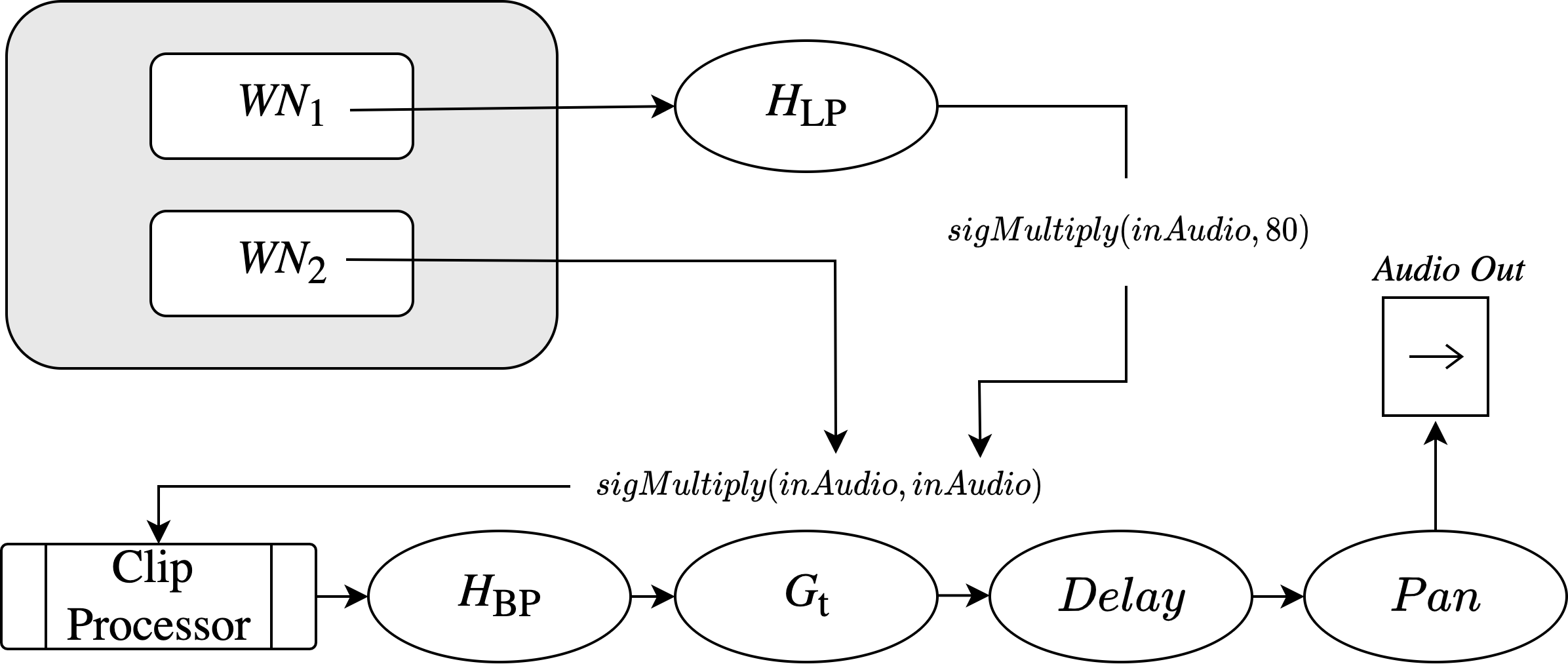}
\caption{Block diagram of the `Afterimage' generator. The base implementation is provided by \cite{fxive}.}
\label{fig:Afterimage}
\end{figure}

The `Afterimage' refers to a second shock audio event, and includes medium-to-long length delays with distant echoing effects \cite{blanco_2008}. It is a direct reaction to the initial thunder \emph{clap} interacting with its surroundings. Fig. \ref{fig:Afterimage} depicts the signal processing graph of this sub-model. Similar to the `Rumbler', the `Afterimage' sub-model processes two independent white noise audio signals $WN_{[1,2]} \sim U[-1,1)$, the contents of which are sampled from a continuous uniform distribution $U$. Intermediate signal $X$ is created as such

\begin{subequations}
    \begin{align}
    &X = (H_{\text{LP}}(WN_1) \cdot 80) \cdot WN_2 \label{eq:afterimageintermediate-a}\\
    &X = \text{max}(\text{min}(X,1),-1) \label{eq:afterimageintermediate-b}\\
    &X = H_{\text{BP}}(X) \label{eq:afterimageintermediate-c}
    \end{align}
\end{subequations}

where $WN_1$ is low-pass filtered by $H_{\text{LP}}$, multiplied by 80 and then by $WN_2$. In a similar fashion to sub-model \ref{ref:themodel-rumbleandroll}, modulating $WN_1$ by $WN_2$ creates a dependency between the signals, emulating the natural relationship between various thunder physics. \eqref{eq:afterimageintermediate-b} demonstrates a clipping of the signal in (-1,1) to ensure stability in that relationship. \eqref{eq:afterimageintermediate-c} shows that intermediate signal $X$ is filtered by $H_{\text{BP}}$, with a pass-band centred at $f=333\text{Hz}$ and $Q=4$. Gain $G_t$ of $X$ ramps periodically through ramping function $R'$ s.t. $R'(G_t)$: $P_{\text{\:initialStrike}} \cdot 2 \xrightarrow[T]{} 0 + \epsilon$, where $d':= d+14s$. Cutoff frequency $f_t$ of $H_{\text{LP}}$ ramps linearly s.t. $R(f_t)$: 33Hz $\xrightarrow[T]{}$ 0Hz. This emphasises the natural attenuation captured in a slowly dissipating thunder sound.

\subsection{Deepener}
\label{ref:themodel-deepener}

The main goal of the `Deepener' is to add magnitude, color, and texture to the lower ends of the frequency content. This sub-model processes a white noise audio signal $WN \sim U[-1,1)$. (\ref{eq:deepenerintermediate}) reflects the signal processing performed on intermediate signal $X$

\begin{subequations}
    \label{eq:deepenerintermediate}
    \begin{align}
    &X = H_{\text{HP}}(H_{\text{LP,1}}(WN)) \cdot 3.5 \label{eq:deepenerintermediate-a}\\
    &X = \text{max}(\text{min}(X,1),-1) \label{eq:deepenerintermediate-b}\\
    &X = H_{\text{LP,2}}(X) \label{eq:deepenerintermediate-c}
    \end{align}
\end{subequations}

in which $WN$ is low-pass filtered by $H_{\text{LP,1}}$, then high-pass filtered by $H_{\text{HP}}$ and then multiplied by 3.5, shown in \eqref{eq:deepenerintermediate-a}. Following this, \eqref{eq:deepenerintermediate-b} denotes the signal clipping in (-1,1), ensuring stable bounds. The signal is then low-pass filtered by $H_{\text{LP},2}$, shown in \eqref{eq:deepenerintermediate-c}. $H_{\text{LP,1}}$ is assigned $f=60$Hz, $H_{\text{LP,2}}$ is assigned $f=80$Hz, and $H_{\text{HP}}$ is assigned $f=15$Hz, and all filters are assigned $Q=3$. Gain $G_t$ of $X$ ramps periodically as such $R'(G_t)$: $P_{\text{growl}} \cdot 6 \xrightarrow[T]{} 0$, where $d':= d+18.5s$.

\begin{figure}[h]
\centering
    \includegraphics[scale=0.093]{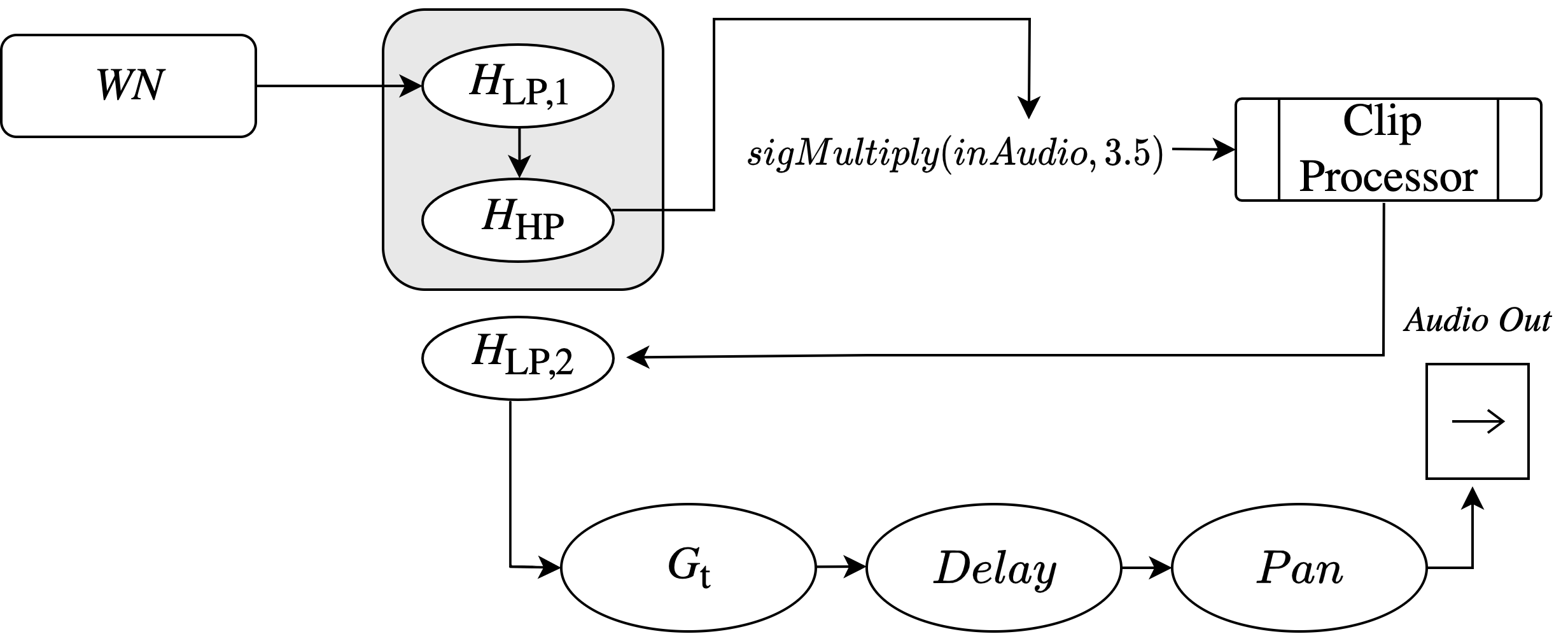}
\caption{Block diagram of the `Deepener' generator. The base implementation is provided by \cite{fxive}.}
\label{fig:deepener}
\end{figure}

\subsection{Post Processing}
\label{ref:postprocessing}

Natural atmospheric occurrences influence the perceived sound of each of these sub-models as they evolve. In order to simulate the impact environmental surroundings have on the generated audio, select post-processing effects are applied to each sub-model. This includes delay, feedback, spatialisation, and convolutional reverb. 

The Web Audio API `DelayNode' is used to create delay and feedback effects by applying a typical delayed sample function \cite{wc3workinggroup}, \cite{reissworking}. A custom `Feedback' node is implemented, parameterised by $P_{\text{delayTime}}$ and $P_{\text{feedback}}$, which represent the delay in seconds until the feedback should begin, and the wet/dry mix, respectively. A delay effect is applied to all sub-models. Feedback is applied to sub-model \ref{ref:themodel-lightningstrikes} s.t. $P_{\text{delayTime}}=0.6$s, $P_{\text{feedback}} = 0.15$. 

The `PannerNode' spatialises the audio in 3-D space, determined by distance, direction, and angle based attenuation. This implementation also defines a head-related transfer function (HRTF). The HRTF convolves the signal with a chosen impulse response based on different angled elevations and the determined attenuation within the three-dimensional space \cite{wc3workinggroup}, \cite{reissworking}. Panning is applied randomly to all four sub-models.

The Web Audio API `ConvolverNode' is used to convolve sub-model \ref{ref:themodel-lightningstrikes} with an impulse response of \cite{warren_2013}, modelling the reverberant acoustics of a natural beach environment \cite{wc3workinggroup}, \cite{reissworking}.

\section{Evaluation}
\label{ref:evaluation}
We conducted a subjective listening test evaluating the perceived realness of our synthesised sound of thunder. The test format was inspired by \cite{tez_selfridge_reiss} which was based on \cite{moffat_reiss_2018}. We asked participants to rate five different audio samples of the sound of thunder on a continuous scale from 1 (extremely unrealistic) to 10 (a real recording). The five audio clips randomly presented to participants included samples of four synthesised models, and one recording of real thunder from the BBC sound archives \cite{bbc}, named `Recording' in Fig. \ref{fig:survey}. Additionally, models from `Farnell' \cite{farnell_2010}, `Fineberg' (our proposed model detailed in Section \ref{ref:themodel}), `FXive' \cite{fxive} and `Saksela' \cite{saksela_2014} were also offered as candidates. For consistency, all samples were played at 44.1kHz.

Using the Absolute Category Rating testing framework offered by Go Listen, participants were prompted to input their ratings after listening to each sample individually \cite{golisten}, \cite{ape2014}.

The participants were a well balanced and unbiased cohort. 54 people, including 26 female-identifying, 26 male-identifying, and 2 gender non-conforming people submitted responses to this survey. The majority of the participants, 48, were not professionals in the field of audio. All participants were between the ages of 17 and 70 with a median age of 30, and a strong majority, 52, reported having normal hearing. The test was conducted remotely and so the listening environment was not controlled. However, a majority of 48 people reported taking the survey in a quiet room. Nine people reported using studio headphones, 15 people used their computer speakers, 16 people listened through consumer headphones, and 13 people used their mobile phone speakers. 

\section{Results}
\label{ref:results}
 \subsection{Quantitative Results}
\label{ref:quantitative-surveyresults}

\begin{figure}[h]
\centering
    \includegraphics[scale=0.262]{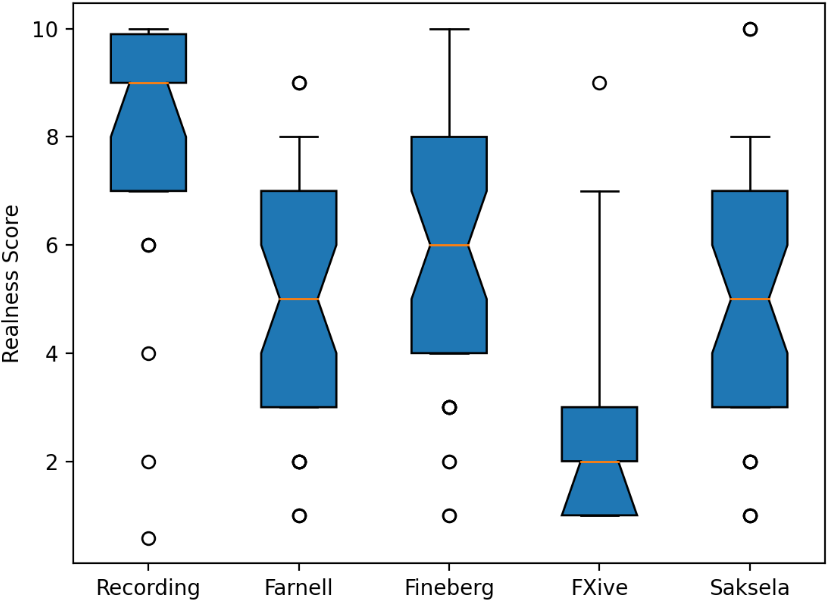}
\caption{Evaluation results for four thunder synthesis models and a real recording.}
\label{fig:survey}
\end{figure}

The quantitative results are depicted in Fig. \ref{fig:survey}. The box plot includes the median rating for each model, indicated by the orange line. The notched box lines represent the 95\% confidence interval around the median. Error bars represent minimum and maximum values and circles represent outliers. The relative perceived realness score of the previously surveyed generative models `Farnell', `FXive', `Saksela' reflect the results reported in \cite{tez_selfridge_reiss}. From these results, it is clear that no synthesised audio sample was perceived as indistinguishable from the real recording. However, the sample created by our model was perceived as more realistic than all other synthesised models surveyed.

\begin{table*}[htbp]
\small
\caption{Select comments on the model described in this paper}
\begin{center}
{\renewcommand{\arraystretch}{1.4}
\begin{tabular}{|c|c|l|}
\hline
\textbf{Participant}&{\textbf{Ranking}}&\textbf{Comment}\\
\hline
A& 7& \pbox{20cm}{\vspace{3pt}``Sounds a bit too movie-like and perfect. I lived on the east coast for long enough and thunder\\was usually more muffled, or much louder''\vspace{3pt}}\\
\hline
B& 3 & \pbox{20cm}{\vspace{3pt}``Sounds synthesised but a reasonable spectrum match. Still didn’t get a sense of it being a\\sound emerging from the distance''\vspace{3pt} }\\
\hline
C& 4 & \pbox{20cm}{``Sounds like thunder but a little too perfect to be real''}\\
\hline
D& 3 & \pbox{20cm}{\vspace{3pt}``Sounds more like a rocket taking off or an explosion than thunder, because of the power and\\pitch of the first sound'' \vspace{3pt}}\\
\hline
E& 6 & \pbox{20cm}{``Sounds close but a single thunderclap is usually faster, and the echo decay is suspicious''}\\
\hline
\end{tabular}}
\label{tab1}
\end{center}
\label{tab:comments-table-eva}
\end{table*}

\begin{table*}[htbp]
\small
\caption{Select comments on model `FXive'}
\begin{center}
{\renewcommand{\arraystretch}{1.4}
\begin{tabular}{|c|c|l|}
\hline
\textbf{Participant}&{\textbf{Ranking}}&\textbf{Comment}\\
\hline
F& 2& \pbox{20cm}{``This sounds like a glitchy fuzzed guitar''}\\
\hline
G& 1 & \pbox{20cm}{``Sounds like a synthesiser, with no low end, no reverb''}\\
\hline
H& 1 & \pbox{20cm}{\vspace{3pt}``If I heard this without the 'sound of thunder' context, I would have thought it was\\ an intro to an 80’s hair band power ballad''\vspace{3pt}}\\
\hline
I& 1 & \pbox{20cm}{\vspace{3pt}``Sounds like a bad video game sound effect, with constant repetitions of the\\ same twanging sound''\vspace{3pt}}\\
\hline
\end{tabular}}
\label{tab1}
\end{center}
\label{tab:comments-table-eva-1}
\end{table*}

\begin{table*}[htbp]
\small
\caption{Scores across hardware}
\begin{center}
{\renewcommand{\arraystretch}{1.4}
\begin{tabular}{|c|c|c|c|c|}
\hline
&\textbf{Consumer Headphones}&\textbf{Studio Headphones}&\textbf{Computer Speakers} & \textbf{Phone Speakers}\\
\hline
Recording& 8.81& 8.38 & 7.8 & 7.66\\
\hline
Farnell& 4.13 & 4.66 & 5.73 & 5.62\\
\hline
Fineberg& 6.5 & 5.88 & 5.46 & 6.69\\
\hline
FXive& 2.69 & 2.22 & 2.33 & 2.53\\
\hline
Sakslea& 5.19 & 5 & 5.13 & 4.62\\
\hline
\end{tabular}}
\label{tab1}
\end{center}
\label{tab:hardware-rankings}
\end{table*}

The real recording of thunder received a median realness score of 8.17. Our model received a median realness score of 6.15 which is higher than `Farnell' whose synthesised sound effect received a median realness score of 5.04, and `FXive' and `Saksela' which received median scores of 2.47 and 5.0, respectively.

\subsection{Qualitative Results}
\label{ref:discussion-surveyresults}
In an attempt to better understand which sonic features exposed the audio as synthesised, the participants were also prompted for open-ended comments on each audio sample. The comments are not used in formal evaluation but rather provide insight into specific characteristics which were perceived as lacking or remarkable.

Select comments summarised in Table \ref{tab:comments-table-eva} highlight common acoustic events which reveal the audio as synthesised. These events were determined to be an unnatural sense of "perfectness", as well as a suspicious initial strike. Responses after hearing the model from \cite{fxive}, the model on which this work was initially based, are summarised in Table \ref{tab:comments-table-eva-1} and are starkly more negative than those in Table \ref{tab:comments-table-eva}.

\section{Discussion}
\label{ref:discussion}
The results detailed in Section \ref{ref:quantitative-surveyresults} confirm the design choices of the proposed model increase the perceived realness of the sound. Based on median value comparison, our model outperformed the other synthesised models by at least 1 ranking score, which is reflective of its enhanced realism. The distribution of our model’s rankings is similar to that of `Farnell' and `Saksela'. This indicates that all three models have the potential to be perceived as realistic, and variance in perception could be explained by individual listening conditions. The `FXive' model's considerably weaker median realness rating and smaller interquartile range reflects a definitively less realistic sounding model, and validates our design choices. Conversely, the relatively small error bars of `Recording', coupled with its near-perfect median rating shows it was, conclusively, the strongest performing thunder sound.

As shown in Fig. \ref{fig:survey} there exist outliers for each surveyed thunder sound source. To the best of our knowledge, the survey participants' age, gender or profession would not influence their ability to evaluate the realism of a recording. However, their geographical location could influence variables such as elevation and humidity, which in turn influence the evolution of pressure disturbances caused by thunder, and so too a listener’s reference for the sound of thunder. In addition to the listening environment, the listening hardware varied across subjects. The frequency response and range required to reproduce the sound of thunder are generally not offered by computer or mobile phone speakers, however as shown in Table \ref{tab:hardware-rankings}, the average rankings across different hardware was negligible.

Comments in Table \ref{tab:comments-table-eva} note that our model sounds too ``perfect'' or ``synthesised''. This is likely explained by the deterministic nature of the signals’ amplitude and cutoff frequency envelope, coupled with uncontrolled distortion. User feedback further critiques our model’s spatialisation technique, which could indicate inferior panning parameters or an unsuitable choice of impulse response for the convolutional reverb.

\subsection{Improvements}
\label{ref:discussion-improvements}

Based on the survey results discussed in Section \ref{ref:discussion-surveyresults}, several modifications were implemented in an attempt to improve the perceived realness of the model. 

Select comment D in Table \ref{tab:comments-table-eva} represents a recurring criticism regarding the pitch, distortion, and power of the \emph{clap} generated by sub-model \ref{ref:themodel-lightningstrikes}. In response, we reduced the cutoff frequency $f_{j,t}$ of the band-pass filter bank $H_{\text{BP}, j}$ in this sub-model by 20Hz. The cutoff frequency $f$ of $H_{\text{HP}}$ in `Deepener' was increased from 15Hz to 30Hz so as to further attenuate inaudible lower frequencies that may have caused unnecessary distortion. To address Table \ref{tab:comments-table-eva} comment A, a Web Audio API dynamic range compressor was appended to the signal chain at unity gain, thresholded at -20dB, with a knee of 20dB, a ratio of 12, and attack and release times of 0 seconds and 0.5 seconds, respectively \cite{wc3workinggroup}. This aided in balancing the explosiveness of the initial `Multi-Strike Lightning' against the following `Rumbler'.

Comments A and C in Table \ref{tab:comments-table-eva} reflect the frequent critique regarding the naturalness of the overall sound. In response, we reduced the resonance of the band-pass filter bank $H_{\text{BP}, j}$ in sub-model \ref{ref:themodel-lightningstrikes} from $Q=10$ to $Q=7$ so that the frequency response alteration is more subtle. We also found that reducing the gain of the overall `Afterimage' from unity to 0.4 aided in reducing the synthesised quality of the tail-end of the sound.

\section{Concluding Remarks}
\label{ref:concludingremarks}

Advances in a physics-inspired, signal-based sound of thunder were implemented. It was demonstrated that our proposed model outperforms all other models evaluated yet is still distinguishable from the real event. In response to user feedback and subsequent synthesis technique analysis, a number of improvements were implemented and future work is identified.

\subsection{Future work}
\label{ref:futurework}

It is worth noting that, while there were no mentions of this in the survey comments, one audio characteristic missing from the event is the crackling noise often emitted as a result of inter-cloud electrical discharge which can emit a sharp and crisp noise right on the brink of the larger thunder \emph{clap}. This addition coupled with a focus on enhancing the sense of natural randomness in the model would achieve a more realistic sound. More specifically the suggestions are as follows:

\begin{itemize}
    \item Simulate the crackling created by electrical discharges directly preceding the main \emph{clap}.
    \item Add more randomness in all sub-models, including gain envelopes and other wave shaping distortions.
    \item Simulate refraction and scattering effects by applying a procedural granular synthesis to sub-model \ref{ref:themodel-lightningstrikes}.
    \item Improve spatialisation parameters for each semantic sub-model of the sound effect.
    \item Conduct a listening test where subjects adjust user input parameters of a model until it sounds the most realistic.
\end{itemize}

The advancement of natural synthesis sound models such as thunder will continue to propel the use of procedural audio in creative industries. This contribution offers more insight into the intricate relationship between procedural audio and nature in the hope that work surrounding this connection will continue to expand the potential of creative and simulated audio spaces.

\bibliographystyle{jaes}

% Reference to bibliography file.
\bibliography{refs}

\end{document}